\pdfoutput=1
\documentclass[12pt]{article}

\usepackage{amsmath,graphicx}
\usepackage{simplewick}
\usepackage{epsf}
\usepackage{graphicx,epsfig}
\usepackage{amsfonts}
\usepackage{amssymb}
\usepackage[nosort]{cite}
\usepackage{setspace}

\def\bk{{\bf k}}

\makeatletter
\renewcommand\section{\@startsection {section}{1}{\z@}%
                                 {-3.5ex \@plus -1ex \@minus -.2ex}%
                                   {2.3ex \@plus.2ex}%
                                   {\normalfont\large\bfseries}}
\renewcommand\subsection{\@startsection{subsection}{2}{\z@}%
                                   {-3.25ex\@plus -1ex \@minus -.2ex}%
                                     {1.5ex \@plus .2ex}%
                                     {\normalfont\bfseries}}
\renewcommand\subsubsection{\@startsection{subsubsection}{3}{\z@}%
                                   {-3.25ex\@plus -1ex \@minus -.2ex}%
                                     {1.5ex \@plus .2ex}%
                                     {\normalfont\itshape}}
\makeatother

\newcommand{\Letter}{
\setlength{\textwidth}{16.5cm}
   \setlength{\textheight}{23cm}
    \hoffset=-0.5in
\voffset=-2.1cm }

\Letter

\renewcommand{\theequation}{\thesection.\arabic{equation}}

\begin{document}
\newcommand{\be}{\begin{equation}}
\newcommand{\ee}{\end{equation}}
\newcommand{\bea}{\begin{eqnarray}}
\newcommand{\eea}{\end{eqnarray}}
\newcommand{\barr}{\begin{array}}
\newcommand{\earr}{\end{array}}
\def\bal#1\eal{\begin{align}#1\end{align}}

\newcommand{\nc}{\newcommand}
\nc{\ba}{\begin{eqnarray}}
\nc{\ea}{\end{eqnarray}}
\newcommand{\calR}{{\cal{R}}}
\newcommand{\calP}{{\cal{P}}}
\newcommand{\calH}{{\cal{H}}}
\newcommand{\calI}{{\cal{I}}}

\renewcommand{\theequation}{\arabic{equation}}

\setcounter{page}{1}
%\begin{flushright}
%\parbox[t]{1.5in}{hep-th/yymmnnn}
%\end{flushright}

\vspace*{0.3in}
\begin{spacing}{1.1}

\begin{center}
{\large \bf A Direct Probe of the Evolutionary History of \\the Primordial Universe}

\vspace*{0.3in} {Xingang Chen$^{1}$, Mohammad Hossein Namjoo$^{1}$, and Yi Wang$^2$}
\\[.3in]
{\em
$^1$ Institute for Theory and Computation, Harvard-Smithsonian Center for Astrophysics,\\
60 Garden Street, Cambridge, MA 02138, USA\\
$^2$Department of Physics, The Hong Kong University of Science and Technology,\\
Clear Water Bay, Kowloon, Hong Kong, P.R.China} \\[0.05in]

\end{center}

\begin{center}
{\bf
Abstract}
\end{center}
\noindent
Since Hubble and Lamaitre's discovery of the expanding universe using galaxies till the recent discovery of the accelerating universe using standard candles, direct measurements of the evolution of the scale factor of the universe $a(t)$ have played central roles in establishing the standard model of cosmology. In this letter, we show that such a measurement may be extended to the primordial universe using massive fields as standard clocks, providing a direct evidence for the scenario responsible for the Big Bang. This is a short and non-technical introduction to the idea of classical and quantum primordial standard clocks.

\vfill

\newpage

%\tableofcontents

%\newpage

Direct measurements of the evolution of the scale factor of the universe $a(t)$ have played central roles in cosmology. By measuring the relation between the distance and redshift of galaxies, Hubble and Lamaitre \cite{Hubble} discovered the expansion of the universe, providing the first evidence for the Big Bang model. Such measurements have been done in exquisite precisions in modern days using the type Ia supernovae as standard candles, leading to the discovery of the late time acceleration of the universe \cite{Riess:1998cb}.
These measurements laid the foundation for the standard model of cosmology.
However, the Big Bang model faces several puzzles regarding its initial conditions, such as the horizon and flatness problems. These problems strongly suggest the existence of a primordial epoch with very different evolutionary history, which gives rise to the Big Bang.

The inflation scenario \cite{Guth:1980zm} has been the leading candidate for the primordial universe. In the mean while, there have been considerable efforts constructing alternative scenarios, such as the string gas, matter bounce and ekpyrotic scenarios \cite{Brandenberger:2016vhg}, each having different $a(t)$ evolution. The alternative scenarios usually suffer from more theoretical problems than inflation, but improvement has been active lines of research.
As usual, the decisive question is how we can observationally distinguish these scenarios.
Following the same footsteps, a natural question is: can we directly measure $a(t)$, the defining property of the primordial universe?

Probing properties of the primordial universe is challenging, simply because the opaque era before the recombination epoch erases most of the information carried by light. So, we rely on a different kind of observables -- the primordial fluctuations -- in which some vital information about the primordial universe has been kept as statistics of large scale distributions of the contents of the universe. Are there any hidden properties in these fluctuations that directly encode $a(t)$?

The primordial gravitational waves (PGW) \cite{Grishchuk:1974ny,Seljak:1996ti}, for a long time, has been the only option. The amplitude of PGW is determined by the gravitational coupling, $H/M_{\rm Planck}$ where $H$ is the Hubble parameter. If one could probe $H$ as a function of horizon-crossing time for different modes, we could probe $a(t)$. For this reason, it is often thought that observing an almost-scale-invariant gravitational wave spectrum implies a constant $H$.
However, two important assumptions have been made, namely, the amplitude of PGW freezes after the horizon crossing (because we observe PGW at the end of the primordial epoch instead of at the mode horizon-crossing), and PGW originates from vacuum fluctuations.
Such assumptions are correct for inflation, but naturally violated in some of the non-inflationary scenarios, such as the matter bounce scenario and string gas cosmology.
For example, in matter contraction \cite{Wands:1998yp}, the amplitude of PGW continues to evolve after the horizon crossing, giving rise to a scale-invariant spectrum even though $H$ is strongly time-dependent.

The degeneracy in using density fluctuations and gravitational waves to probe $a(t)$ is not a coincidence. Fluctuations at different length scales are generated at different physical times, but we observe all modes at one instant. The degeneracy would be broken if there were a standard clock that can label the passage of time on those fluctuations as they are being generated.

Fortunately, this is indeed the case and in this letter we shall introduce the primordial standard clock (PSC). There exist many massive fields in the primordial universe. These massive fields originate from the UV completion (such as the moduli fields, KK modes and stringy excitations, which are typically quite heavy) and IR uplifting (such as the standard model fields and other light fields, which become heavier through radiative corrections or couplings to the background curvature).
If the massive fields are heavy enough, at some point during their evolution in any time-dependent background, they oscillate either classically or quantum mechanically in a model and scenario-independent way, similarly to that of harmonic oscillators in the flat spacetime.
These oscillations can be regarded as a PSC, generating ticks for the time coordinate. The ticks get imprinted in the density fluctuations in terms of special oscillatory features -- the clock signals -- that directly encode $a(t)$ \cite{Chen:2011zf,Chen:2014cwa,Chen:2015lza,Chen:2016cbe}.

To demonstrate how $a(t)$ is encoded in the clock signal, let us first look at the case of the classical PSC \cite{Chen:2011zf,Chen:2014cwa}. In this case, the oscillation of massive fields is excited by certain sharp features in models. Once excited, the massive field $\sigma$ oscillates as a background component if the mass $m$ is larger than the horizon scale,
\bea
\sigma \propto e^{imt} ~.
\label{eq:massive}
\eea
These oscillations induce small oscillatory components to the couplings in various correlation functions.
On the other hand, the density fluctuation $\zeta$ corresponds to a massless mode in the primordial universe. At subhorizon, this mode oscillates as
\begin{align}
  \label{eq:massless}
  \zeta_\mathbf{k} \propto e^{-i k \tau}~,
\end{align}
where the conformal time $\tau$ is related to the physical time $t$ by $d\tau = dt / a(t)$, and $\mathbf{k}$ is the comoving momentum of the mode.

\begin{figure}[t]
  \centering
  \includegraphics[width=0.8\textwidth]{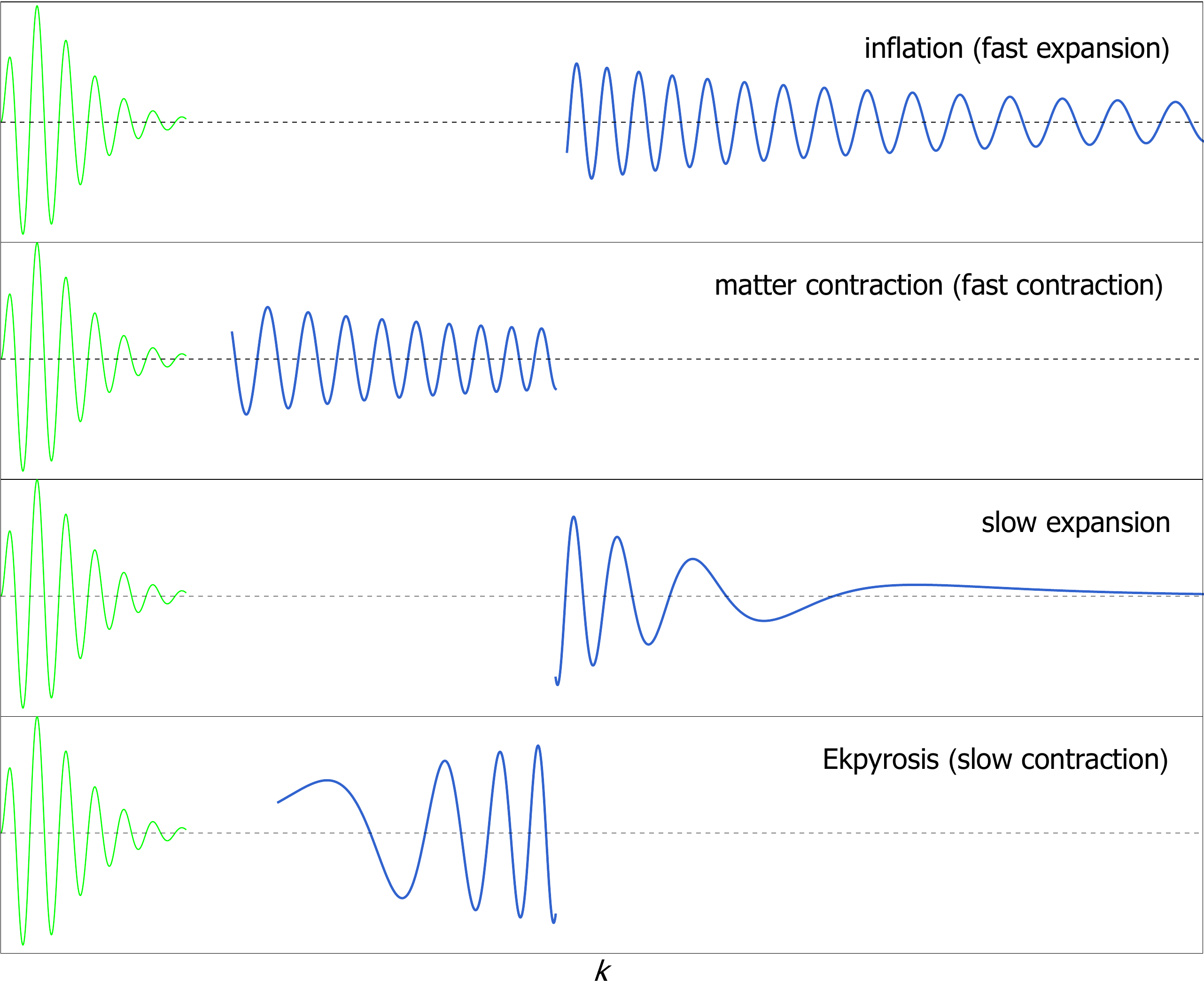}
  \caption{\label{fig:scenarios2} \small The classical PSC signals, plotted as the amplitude of the oscillatory component of the two-point correlation function as a function of the scale $k$, for four different types of scenarios characterized by different $a(t)$. The light green lines are the sharp feature signals, which are qualitatively the same for different scenarios. The dark blue lines are the clock signals. The relative spacing between the ticks of the clock signals, namely the phases of these oscillations, directly encodes $a(t)$.
  Some less model-independent properties such as the envelops of the signals are also shown in Fig.\ref{fig:scenarios2} \& \ref{fig:qscenarios}.}
\end{figure}

\begin{figure}[t]
  \centering
  \includegraphics[width=0.8\textwidth]{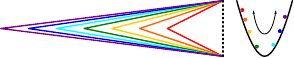}
  \caption{\label{fig:squeezed} \small Squeezed configurations of a three-point function of momenta. In the left panel, the dashed black line denotes the longer mode (i.e.~with smaller momentum) -- the fluctuation mode from the massive field (which is later converted to density fluctuation). The lines colored from purple to red denote two short modes (i.e.~with larger momenta) -- the density fluctuation modes, with their physical wavelengths from short to long. Those modes are in resonance with the massive mode when the massive mode has different phases of quantum oscillation (right panel).}
\end{figure}

\begin{figure}[h]
  \centering
  \includegraphics[width=0.8\textwidth]{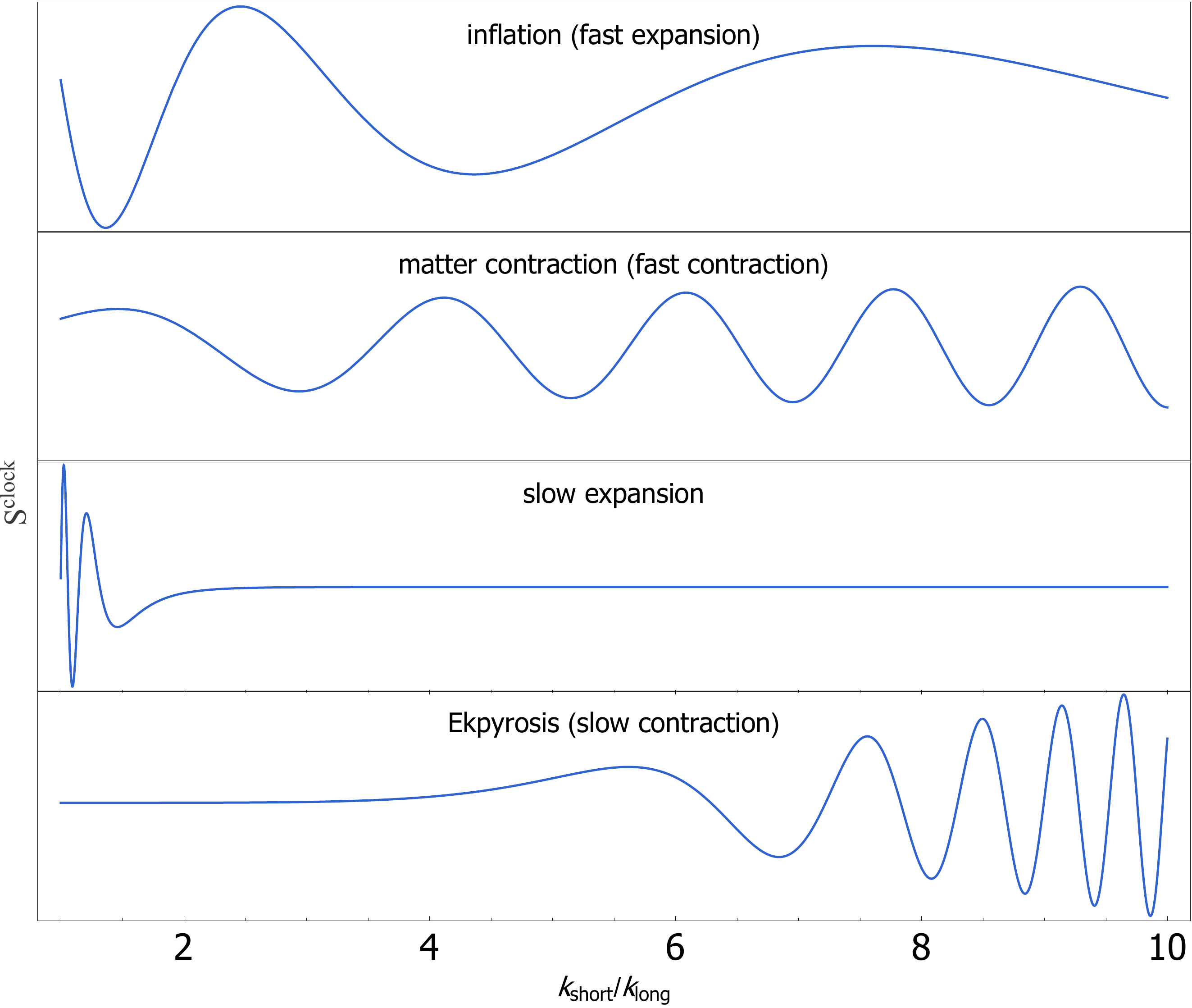}
  \caption{\label{fig:qscenarios}\small The quantum PSC signals, plotted as the amplitude of the three-point function $S^{\rm clock}$ as a function of the momentum ratio $k_{\rm short}/k_{\rm long}$, for four different types of scenarios characterized by different $a(t)$.}
\end{figure}

The correlation function between $\zeta$'s schematically contains the following term,
\bea
\langle \zeta_{\bk}^2\rangle
\supset
  \int e^{i (mt -2 k \tau)} d\tau~.
\label{eq:resonance-integral}
\eea
Notice that the oscillation frequency of $\zeta$ is $t$-dependent, given by the background evolution. Although at most places this integral averages to zero, once the frequency of $\zeta$ matches to that of the massive clock field, the resonance condition is satisfied
\begin{align}
  \label{eq:resonance-condition}
  \frac{d}{dt} \left ( mt -2k\tau \right ) =0 ~,
\end{align}
and the correlation function receives a large contribution \cite{Chen:2008wn}
\bea
\langle \zeta_{\bk}^2\rangle
\supset
 e^{i (mt_* -2 k \tau_*)} ~,
\eea
where $t_*$ or $\tau_*$ denotes the time at which the resonance happens,
\bea
a(t_*)=a(\tau_*) = 2k/m ~.
\eea
Using these relations we see that
\bea
\langle \zeta_{\bk}^2\rangle
\supset
 \exp\left[i m~t(2k/m) -2i k~\tau(2k/m) \right] ~,
\label{2pt_resonance}
\eea
where $t(2k/m)$ and $\tau(2k/m)$ are inverse functions of the scale factor $a(t)$ and $a(\tau)$, respectively.
Therefore, the scale factor evolution is directly recorded in the phase of this signal as a function of the comoving momentum $k$.
Superhorizon physics can only change the envelops of the clock signals.
See Fig.~\ref{fig:scenarios2} for illustrations.

The classical PSC requires presence of sharp features, which limits its generality. So it still remained a question whether there exist some types of signals that are as general as PGW, and at the same time can be used to model-independently distinguish different primordial universe scenarios.
It has been realized recently \cite{Chen:2015lza,Chen:2016cbe} that the answer to this question is affirmative! Massive fields quantum-fluctuate in any time-dependent background, and if their masses are large enough these fluctuations behave very much like the classical oscillations \eqref{eq:massive}. As we will show, these oscillations can be just as well used as PSC. Since there are plenty of massive fields in any realistic primordial universe model, and they couple to the density fluctuation field at least gravitationally, the clock signals also exist ubiquitously in any model and scenario.

An immediate difference between the classical and quantum PSC is the following: in the classical case, the sharp feature is a source of the scale-invariance breaking, which is why the classical clock signals are various scale-dependent oscillatory features; while in the quantum case, the quantum fluctuation itself is not a source of the scale-invariance breaking, so the clock signals must show up in a very different manner. A detailed investigation confirms this expectation \cite{Chen:2015lza}, and the results can be explained concisely as follows.

The key is to look at the correlation functions that are of higher points than two-point (namely, the non-Gaussianities), and look for correlations that involve massive fields as mediators. The simplest case, which nonetheless illustrates the key points, is the three-point function depicted in Fig.~\ref{fig:squeezed}, in which one massive mode (which later converts to density fluctuation) couples to two density fluctuation modes.

Let us look at the squeezed configuration in which the massive field is the long mode. In such a configuration, the massive field basically serves as a uniform oscillating background for the two short curvature modes.
This is now a familiar situation that we have encountered in the classical PSC case.
The correlation function is schematically
\begin{align}
  \label{eq:3pt}
  \left \langle \zeta_ \mathbf{k_{\rm short}}^2 \zeta_{\bk_{\rm long}} \right \rangle
  \propto
  \left \langle \zeta_ \mathbf{k_{\rm short}}^2 \sigma_{\bk_{\rm long}} \right \rangle
  \supset
  \int e^{i (mt -2 k_{\rm short} \tau)} d\tau~.
\end{align}
The rest of the reasoning is very similar to the classical case following \eqref{eq:resonance-integral}, except that here we have required that $k_{\rm short}$ be large {\em relative} to $k_{\rm long}$. So in the final result \cite{Chen:2015lza}, the dependence on the scale $k$ in the classical PSC case is replaced by the dependence on the ratio of the scales, $k_{\rm short}/k_{\rm long}$, namely the shape of the momentum triangle.
{\em The quantum PSC signals manifest as shape-dependent oscillatory features in non-Gaussianities.} This is illustrated in Fig.~\ref{fig:qscenarios}. Unlike the classical clock case in which the mass of the clock fields can be much larger than the horizon scales, the most interesting quantum clock fields are those with mass slightly heavier, but still of order the horizon scales, because heavier ones would have to pay a price of a Boltzmann suppression factor \cite{Arkani-Hamed:2015bza}. Interestingly, at least for inflation, such fields are generically expected in supergravity model building \cite{Copeland:1994vg,Baumann:2011nk}.

To look for the clock signals in density fluctuations, we search for the scale-dependent classical PSC signals in all correlation functions, and the shape-dependent quantum PSC signals in non-Gaussianities. The analyses apply to all manifestations of the primordial density fluctuations, including the cosmic microwave background, large scale structures, and 21cm tomography of the cosmic hydrogen atoms.

For inflation, it has been shown that the mass and spin spectra of the theory are encoded in the squeezed limit of primordial non-Gaussianities \cite{Chen:2009we,Arkani-Hamed:2015bza,Assassi:2012zq,Noumi:2012vr}.
Much like particle physics, detecting particles and measuring their properties rely on resonances. Thus, by looking for the PSC signals, we are also searching for new particles from a cosmological collider that nature has built and run for us \cite{Arkani-Hamed:2015bza}.

\medskip
\section*{Acknowledgments}
XC and MHN are supported in part by the NSF grant PHY-1417421.
YW is supported by Grant HKUST4/CRF/13G issued by the Research Grants Council (RGC) of Hong Kong.

\end{spacing}

%\newpage

\end{document}